\documentclass{article}
\usepackage[dvips]{graphicx}

\title
{
GeV-TeV Gamma-ray Astronomy
}

\author{Masaki Mori \\ \small
Institute for Cosmic Ray Research, University of Tokyo, Kashiwa, Chiba 277-8582, Japan }
\date{}

\begin{document}
\maketitle 

\begin{quotation}
Recent results of GeV and TeV observations of gamma-rays
from the Universe are briefly reviewed. Topics include observational technique, diffuse gamma-rays, pulsars, unidentified sources, plerions, supernova remnants and AGNs. 
\end{quotation}
\sloppy
\maketitle

\section{Introduction}

The history of gamma-ray astronomy began in the 1950's
when Hayakawa \cite{Hay52} and Morrison \cite{Mor58}
predicted that gamma-rays would be produced via neutral pion decays generated in collisions of cosmic-rays with interstellar matter. Cherenkov observation of air showers was pioneered by Jelly \cite{Gal53} and Chudakov applied this technique to search for gamma-ray signals from celestial sources \cite{Chu65}. Since this early time the Crab nebula was the most promising source due to its high activity \cite{Coc59, Gou65}. Point sources were found by satellites such as SAS-II and COS B in the 1970's and early 1980's, but the launch of the Compton gamma-ray observatory was the beginning of a new era with its discovery of more than 200 sources \cite{Fic97, Geh00}. On the other hand, ground-based Cherenkov telescopes with imaging capability started to produce reliable detections of sources in the late 1980's \cite{Wee96}.

A distinctive feature of gamma-ray astronomy is the inherent non-thermal origin of radiation at these energies, regardless of its detailed mechanism such as bremsstrahlung or inverse Compton scattering of high-energy electrons in matter or in radiation fields and decay of neutral pions generated in collisions of high-energy protons in interstellar matter. In other words, gamma-ray astronomy can reveal violent aspects of the Universe, far apart from the thermal Universe where most processes are in equilibrium.

\section{GeV observations}

The most complete data up to now in the GeV region were provided by the EGRET detector onboard the Compton gamma ray observatory launched in 1991 and returned to the Earth in 2000 \cite{Tho93}. It was a combination of a spark chamber tracker and a scintillation calorimeter to catch pair production processes caused by gamma-rays and covered the energy range from 30~MeV to 30~GeV. The angular resolution depended on energy but was about two degrees at 1~GeV for each event.

\begin{figure} 
\begin{center}
\includegraphics[height=4cm]{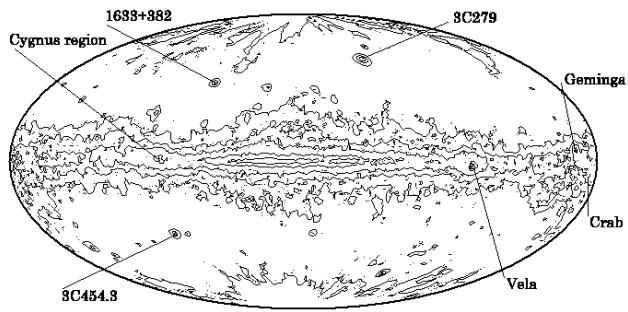}
\caption{Contour map of gamma-ray intensities above 100~MeV measured by EGRET  in the galactic coordinates \protect\cite{Stanford}. Some sources are marked.} 
\label{fig:contmap} 
\end{center}
\end{figure}

Fig.\ref{fig:contmap} shows the intensity map of gamma-rays above 100~MeV observed by EGRET plotted in galactic coordinates \cite{Stanford}. Obviously the galactic plane is the strongest source of gamma-rays in this energy region. %

Strong point sources such as Vela pulsar, Crab pulsar/nebula and Geminga can be identified easily, but weaker sources, especially on the galactic plane, can be identified only after subtraction of the galactic diffuse emission.

Modeling the diffuse emission is a complicated matter
but the EGRET instrument team succeeded in construction of
a detailed model which describes the profile of emission fairly well based on three dimensional modeling of the matter distribution and cosmic ray intensity \cite{Ber93, Hun97}.

Although the general spatial distribution of diffuse gamma-rays fits the data well, the energy spectrum poses a question: one can see the observed spectrum above about one GeV is about 50 percent higher than calculation \cite{Hun97}. In this energy region, nuclear gamma-rays from neutral pion decay is believed to be dominant, but some non-standard parameters may have to be introduced to explain this excess. Such models include a flatter proton spectrum \cite{Mor97,Voe00}, flatter electron spectrum \cite{Por97, Poh98, Str00} and so on. Atmospheric Cherenkov observations and balloon observations impose some limits on possible models \cite{Hun01}.

If we subtract the diffuse model from the observed distribution, point sources can be identified as local peaks. In the real analysis, peaks and the diffuse model are separated using a maximum likelihood procedure, allowing absolute normalization ambiguity of the model fluxes \cite{Mat96}.

\begin{figure} 
\begin{center}
\includegraphics[height=6.5cm]{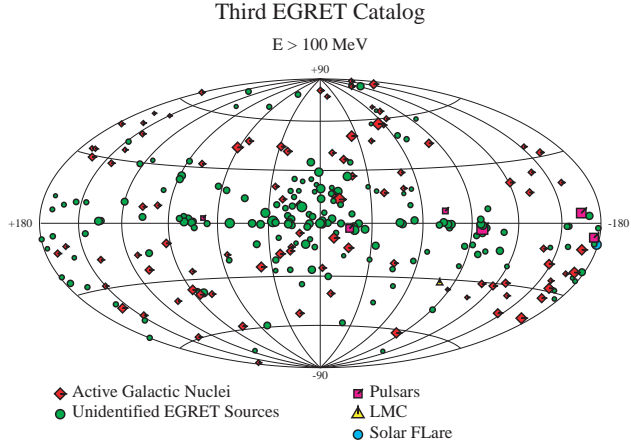}
\caption{Plot of source locations in the 3rd EGRET catalog \protect\cite{Har99}.} \label{fig:3EG} 
\end{center}
\end{figure}

Fig.\ref{fig:3EG} shows the well known map of point sources detected by EGRET \cite{Har99}. 
\begin{table} 
\begin{center}
\caption{Summary of the 3rd EGRET catalog \protect\cite{Har99}.} 
\label{tab:3EG} 
\begin{tabular}{cc} \hline Category & Number of sources \\ \hline Pulsars & 5 \\ AGN (mostly blazars) & 66 + 27 (marginal) \\ Radio galaxy (Cen A) & 1 (marginal) \\ Unidentified (some may be SNRs) & 170 \\ Large Magellanic Cloud & 1 \\ Solar flare & 1 \\ \hline 
\end{tabular} 
\end{center}
\end{table}

One can see most of sources near the galactic plane are unidentified due to EGRET's lack of angular resolution. Table~\ref{tab:3EG} is the summary of source classes in the EGRET catalog \cite{Har99}.

\subsection{Pulsars}
\begin{figure} 
\begin{center}
\includegraphics[height=7cm]{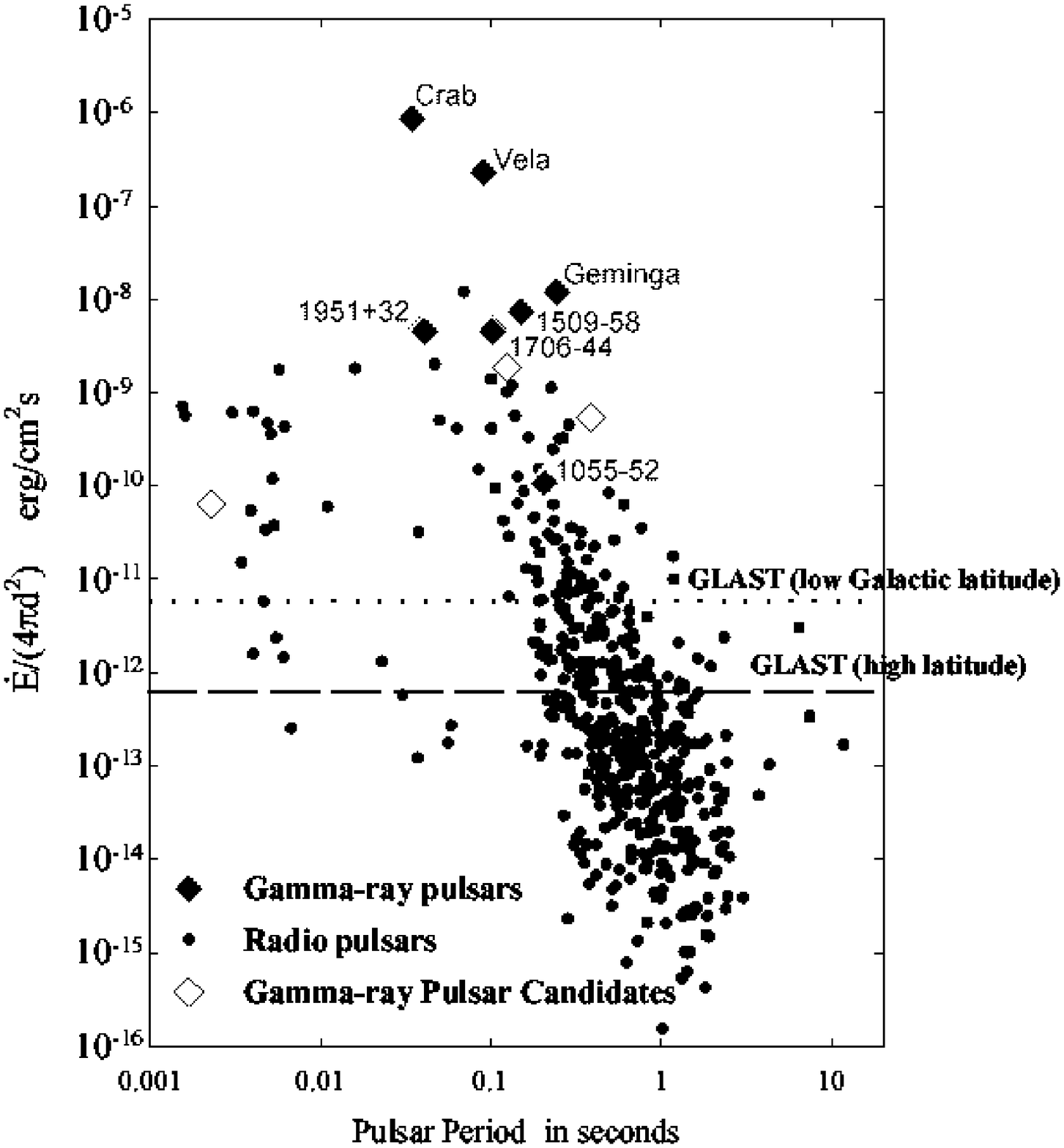}
\caption{Rotation energy loss of pulsars plotted against pulsar period  \protect\cite{Tho00}. Gamma-ray pulsars are marked.} 
\label{fig:pulsar} 
\end{center}
\end{figure}

Seven pulsars have been identified in gamma-rays due to their pulse periods, which coincide with those at other wavelengths \cite{Tho00}. In Fig.\ref{fig:pulsar}, pulse period is plotted against rotation energy loss for known pulsars \cite{Tho00}. The statistics are too small to draw any conclusion about the distribution, but most of gamma-ray pulsars are in the top ranks in rotation energy fluxes, $\dot{E_{\rm rot}}/(4\pi d^2)$ where $d$ is the distance to the pulsar. There are some other gamma-ray pulsar candidates but their identification is marginal \cite{Tho01}.%

Light curves at various wavelengths can provide some hint of the pulse emission mechanisms. Some show double-peaked structure. The relative phases of the peaks are different at different wavelengths, except for the Crab. (See the reference \cite{Tho00} for details.)

\subsection{Active Galactic Nuclei}
\begin{figure} 
\begin{center}
\includegraphics[height=5cm]{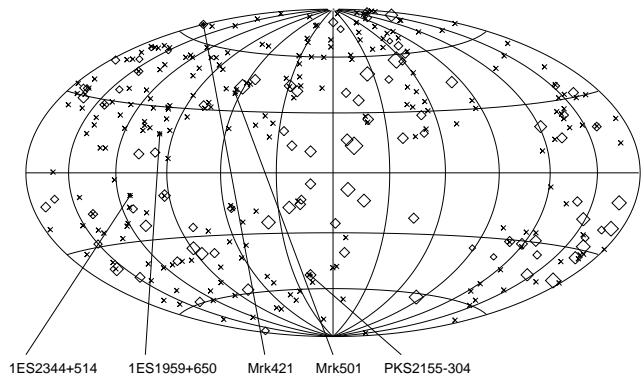}
\caption{Map of 233 BL Lac objects based on catalog by Padovani and Giommi \protect\cite{Pad95} (crosses) and EGRET detected blazars (diamonds). TeV blazars are also specified.} 
\label{fig:bllac} 
\end{center}
\end{figure}

Most of EGRET-detected AGNs are blazars
with jets that are believed to be pointing close to the line-of-sight to our galaxy \cite{Muk01}. Fig.\ref{fig:bllac} is a plot of these with BL~Lac objects compiled by Padovani and Giommi \cite{Pad95}. The list of BL Lacs is incomplete and the southern hemisphere must be explored more. %

The redshift distribution of EGRET-detected blazars extends
to $z\sim2$, but those high redshift blazars are not
promising TeV emitters as will be explained later.
The wide-band spectral energy distributions are studied
for several BL Lac objects.
They show a double peak structure which is ascribed to synchrotron emission and inverse Compton emission of high energy electrons \cite{Fos98}. Blazars with synchrotron peaks at higher frequencies tend to have flatter spectra and can emit higher energy photons \cite{Lin99}, which is believed to be the case of TeV blazars as will be mentioned later.

Time variability of fluxes at various wavelengths has been studied for some sources, especially 3C279 \cite{Har01}, to limit models of particle acceleration in jets.

\subsection{Unidentified sources}
Most EGRET sources have not been identified with counterparts at 
other wavelengths, since the angular resolution of EGRET is limited. Grenier divided them into two classes, persistent and non-persistent \cite{Gre01}. Persistent sources are distributed at low latitudes and can be associated with Geminga-like pulsars, supernova remnants OB associations or Gould belt. Non-persistent sources are rather far from the galactic plane and can be associated with the galactic halo \cite{Gre01}.

\begin{figure} 
\begin{center}
\includegraphics[height=5cm]{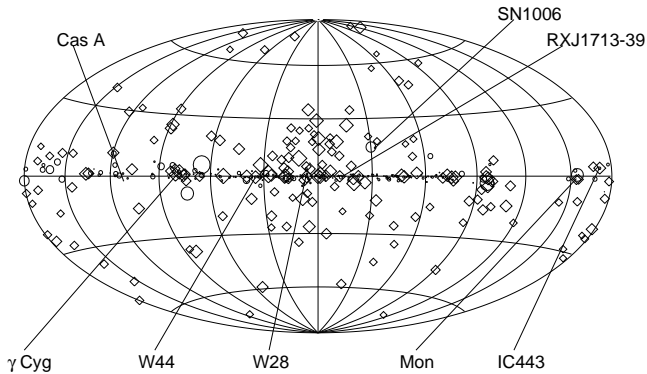}
\caption{Map of SNRs based on the catalog by Green \protect\cite{Gre00}
(circles) and EGRET unidentified objects (diamonds).
Possible identified SNRs by Esposito et al. \protect\cite{Esp96} (bottom) and TeV sources (top) are marked .} 
\label{fig:snrs} 
\end{center}
\end{figure}

Fig.\ref{fig:snrs} is an example plot showing the correlation of EGRET unidentified sources with supernova remnants \cite{Gre00}. Some can be associated \cite{Esp96}, but supernova remnants cannot be a major population of unidentified sources. 

\subsection{Extragalatic diffuse emission}
Another important observation by EGRET is the existence of a uniform gamma-ray background, which must have extragalactic origin. It extends beyond 100~GeV with a single power-law, $E^{-2.10\pm0.03}$ \cite{Sre98}. One natural explanation is a superposition of unresolved point sources like blazars, but some authors estimate that blazars alone are insufficient to explain this flux 
\cite{Ste96, Mue00, Muk99}.

\subsection{Gamma ray bursts}
Five gamma-ray bursts were detected by EGRET \cite{Din95}. Their average spectrum, $E^{-1.95\pm0.25}$, is flat and may extend to higher energies, which can give us hints on their origin \cite{Man96}.

\section{TeV observations}

Ground-based imaging Cherenkov telescopes are becoming
a powerful tool to study
very high energy gamma-rays with their capability to discriminate gamma-rays from background protons \cite{Wee97}.

\begin{figure} 
\begin{center}
\includegraphics[height=5cm]{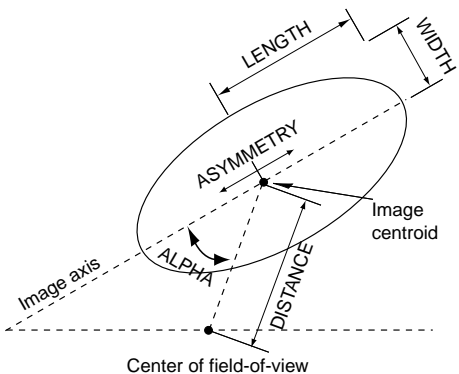}
\caption{Definitions of image parameters.} 
\label{fig:imagepara} 
\end{center}
\end{figure}

\begin{figure} 
\begin{center}
\includegraphics[height=7.5cm]{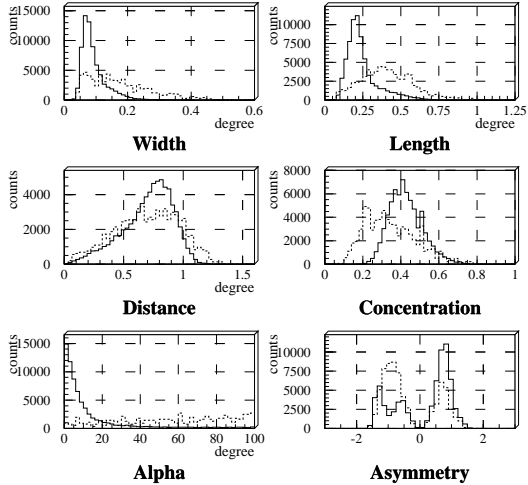}
\caption{Typical distribution of image parameters obtained
by Monte Carlo simulations.
Solid histograms are for gamma-rays and dotted ones for protons.} 
\label{fig:imageparadistr} 
\end{center}
\end{figure}

Gamma-ray images come from pure electromagnetic showers and
are sharp and oriented toward the object being tracked.
They can be separated from nuclear cosmic-ray showers using imaging parameters \cite{Hil85}: width, length, distance, alpha, asymmetry (Fig.\ref{fig:imagepara}). Fig.\ref{fig:imageparadistr} shows the difference in imaging parameter distributions for gamma-rays and protons obtained by Monte Carlo simulations. Among these, alpha is the most powerful discriminator of all for single telescope combined with other parameters to cut non-gamma-ray events from data \cite{Pun91}. Usually candidate objects are observed using ON- and OFF-source modes, and the gamma-ray signal is extracted from subtraction of the OFF-source alpha distribution from that of the ON-source. %

\begin{figure} 
\begin{center}
\includegraphics[height=4.5cm]{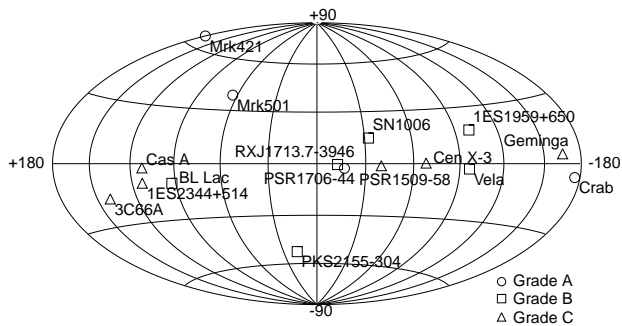}
\caption{Plot of TeV gamma-ray sources  listed in Table \protect\ref{tab:tevcat} in the Galactic coordinates.} 
\label{fig:tevcat} 
\end{center}
\end{figure}

Table \ref{tab:tevcat} is a catalog of TeV gamma-ray objects classified by Weekes \cite{Wee01}, which are plotted in galactic coordinates in Fig.\ref{fig:tevcat}. There are four established sources (Grade~A: a $5\sigma$ detection with an equally significant verification by another experiment). Two of them are galactic sources related to pulsar nebulae and others are extragalactic BL Lac objects. Other sources (Grade~B: a $5\sigma$ detection but without independent verification, Grade~C: a strong detection but with some qualifications) require confirmation. 
\begin{table} 
\caption{TeV gamma-ray source catalog by Weekes \protect\cite{Wee01}.} 
\label{tab:tevcat} 
\begin{center} 
\begin{tabular}{llllll} \hline 
\small 
Source & Type & $z$ & {Discovery}  & {EGRET} & {Grade} \\ \hline 
\multicolumn{3}{l}{\bf Galactic sources} & & & \\ 
 Crab Nebula & Plerion & & 1989 & yes & A \\ 
 PSR 1706$-$44 & Plerion? & & 1995 & no & A \\ 
 Vela & Plerion? & & 1997 & no & B \\  SN1006 & Shell & & 1997 & no & B$-$ \\  
 RXJ1713.7$-$3946 & Shell & & 1999 & no & B \\  
 Cassiopeia A & Shell & & 1999 & no & C \\  
 Centaurus X-3 & Binary & & 1999 & yes & C \\ 
\multicolumn{3}{l}{\bf Extragalactic sources} & & & \\ 
 Markarian 421 & XBL & 0.031 & 1992 & yes & A \\  
 Markarian 501 & XBL & 0.034 & 1995 & yes & A \\  
 1ES 2344+514 & XBL & 0.044 & 1997 & no & C\\  
 1ES1959+650 & XBL & 0.048 & 1999 & no & B$-$\\ 
  PKS 2155$-$304 & XBL & 0.116 & 1999 & yes & B\\ 
  3C66A & RBL & 0.44 & 1998 & yes & C\\ \hline 
\%end{fulltabular} 
\end{tabular} 
\end{center} 
\end{table} 

\subsection{Plerions}
Table \ref{tab:plerion} is a summary of TeV observations of plerions (or pulsar nebula) by Fegan \cite{Feg00}. 
\begin{table} 
\begin{center}
\caption{TeV observation of Plerions \protect\cite{Feg00}.} \label{tab:plerion} 
\begin{tabular}{lcc} \hline 
Object & Exposure time & Flux/Upper limit \\
Name   & (hours)& $\times 10^{-11}$ cm$^{-2}$s$^{-1}$ \\ \hline
{\bf EVERYONE} &     &     \\
Crab Nebula    & $\rightarrow \infty$ & 7.0 ($>400$ GeV) \\
{\bf CANGAROO} &   &  \\
\,\,\,PSR 1706$-$44 &  60 & $0.15$ ($>1$ TeV) \\
\,\,\,Vela Pulsar &  116 & $0.26(E/2{\rm TeV}^{-2.4})$ TeV$^{-1}$ \\
{\bf Durham} &   &  \\
\,\,\,PSR 1706$-$44 &  10 & $1.2$ ($>300$ GeV) \\
\,\,\,Vela Pulsar &  8.75 & $<5.0$ ($>300$ GeV) \\ 
\end{tabular} 
\end{center}
\end{table}

The wide-range spectral energy distribution of unpulsed emission from the Crab nebula shows two peaks which are interpreted as synchrotron emission from high energy electrons and inverse Compton scattering of synchrotron photon by the same electrons \cite{Jag96, Aha98}. The gap between satellite and ground-based experiments is now being filled by Cherenkov telescopes using large-area solar power collectors, namely STACEE \cite{Ose01} and CELESTE \cite{Nau01}.

The pulsed spectrum needs further study. There is no report of the detection of the pulsed component at TeV energies. The spectral break in this energy region is a good indicator of the particle acceleration site around pulsars and we may discriminate between polar cap models, outer gap models and others \cite{Har01a}.

\subsection{Supernova remnants}
Supernova remnants are long considered as acceleration sites of cosmic rays, although direct evidence has been lacking. Recently several detections at TeV 
energies have been reported
(Table \ref{tab:SNR}).
\begin{table}
\begin{center}
\caption{TeV observations of shell-type supernova remnants \protect\cite{Feg00}.} \label{tab:SNR} \begin{tabular}{@{\hspace{\tabcolsep}\extracolsep{\fill}}lcc} \hline Object & Exposure time & Flux/Upper limit \\
Name   & (hours)& $\times 10^{-11}$ cm$^{-2}$s$^{-1}$ \\ \hline
{\bf CANGAROO} &   &  \\
\,\,\,RXJ 1713.7$-$3946 &  66 & $0.53$ ($>1.8$ TeV) \\ \,\,\,SN1006  &  34 & $0.46$ ($>1.7$ TeV) \\ \,\,\,W28  &  58 & $<0.88$ ($>1.5$ TeV) \\
{\bf HEGRA} &   &  \\
\,\,\,Cas A &  232 & $0.058$ ($>1$ TeV) \\
\,\,\,$\gamma$ Cyg &  47 & $<1.1$ ($>500$ GeV) \\
{\bf Durham} &   &  \\
\,\,\,SN1006  &  41 & $<1.7$ ($>300$ GeV) \\
{\bf Whipple} &   &  \\
\,\,\,Monoceros  &  13.1 & $<4.8$ ($>500$ GeV) \\
\,\,\,Cas A  &  6.9 & $<0.66$ ($>500$ GeV) \\
\,\,\,W 44  &  6 & $<3.0$ ($>300$ GeV) \\
\,\,\,W 51  &  13.1 & $<3.6$ ($>300$ GeV) \\
\,\,\,$\gamma$ Cyg &  9.3 & $<2.2$ ($>300$ GeV) \\
\,\,\,W 63  &  2.3 & $<6.4$ ($>300$ GeV) \\
\,\,\,Tycho  &  14.5 & $<0.8$ ($>300$ GeV) \\
{\bf CAT} &   &  \\
\,\,\,Cas A  &  24.4 & $<0.74$ ($>400$ GeV) \\
\end{tabular}
\end{center}
\end{table}

The supernova remnant 1006 shows shell structure and lacks a central energy source. Non-thermal X-ray emission near the rim, detected by ASCA \cite{Koy95}, indicates the existence of high energy electrons up to 100~TeV. CANGAROO detected a TeV signal whose peak is consistent with the northeast rim \cite{Tan98}.

The TeV emission can be explained by inverse Compton scattering of microwave background photons by high energy electrons \cite{Nai99}. The data is fitted well if we assume a magnetic field strength of $\sim4$\,$\mu$G. If we assume the neutral pion decay process instead, upper limits imposed by EGRET make fits difficult. Thus there is no evidence of proton acceleration here, giving no hint to the long-standing problem of cosmic-ray acceleration in supernova remnants.

The recent detection of two supernova remnants at TeV energies, RXJ~1713.7$-$3946 (G347.3$-$0.5) \cite{Mur00} and Cassiopeia~A \cite{Pue01}, both of them lacking central engines, can provide new keys to the problem.

\subsection{Active galactic nuclei}
Table \ref{tab:blazar} is a summary of TeV observations of active galactic nuclei \cite{Kre00}. Fluxes at TeV energies are more variable than at other wavelengths and often repeated detection is difficult. 
\begin{table} 
\caption{TeV observations of active galactic nuclei \protect\cite{Kre00, Hor01}.} \label{tab:blazar} 
\begin{center} 
\begin{tabular}{lcclc} \hline
Source & Energy & Flux      & Group & EGRET \\
       & (GeV)  &           &       & source \\ \hline
\multicolumn{2}{l}{\underbar{\bf Blazars: XBL}} &   &  &  \\
Markarian 421 & 260 & variable & Whipple,  & yes \\ \,\,\,$z=0.031$ & & & HEGRA, CAT  & \\ Markarian 501 & 260 & variable & Whipple,  & no \\ \,\,\,$z=0.034$ & & & HEGRA, CAT, & \\  & & &  TA \cite{Ame00}& \\
1ES2344+514 & 300 & variable & Whipple & no \\
\,\,\,$z=0.044$ & & & & \\
PKS2155$-$304 & 300 & variable & Durham & yes \\ \,\,\,$z=0.116$ & & & & \\
1ES1959+650 & 600 & variable & TA & no \\
\,\,\,$z=0.048$ & & & & \\
\multicolumn{2}{l}{\underbar{\bf Blazars: RBL}} &    &  &  \\
3C66A & 900 & variable & Crimea & yes \\
\,\,\,$z=0.44$ & & & & \\
\end{tabular}
\end{center}
\end{table}

The variability of Mrk~421 is very fast and the doubling time can be less than one hour \cite{Gai96}. Multiwavelength campaigns place constraints on the location and size of the emission region \cite{Tak00}.

The spectrum again shows a double-peaked structure which is considered as synchrotron emission of high energy electrons and inverse Compton scattering of synchrotron photons by those electrons \cite{Bic01}. This model gives us parameters such as beaming factor of around 10 and magnetic field of around 0.1~G. However, proton acceleration models are still applicable, as has often been discussed \cite{Sik01}.

The problem here is the theoretical prediction of the attenuation of TeV photons in the intergalactic space via the production of electron-positron pairs in collisions with infrared background photons \cite{Ste92}. There is a large uncertainty in the flux of infrared photons since observation is quite difficult, but in most cases TeV photons cannot go beyond several tens of megaparsecs, corresponding to redshifts greater than 0.1 \cite{Pri01}.

The observed spectrum of Mrk501 extends to around 10~TeV 
and shows some steepening \cite{Aha01}.
If it is corrected for attenuation, the original spectrum flattens above 10~TeV, which is hard to understand \cite{Pro00}. This means the infrared intensity is overestimated or, as some authors have suggested, the Lorentz invariance is violated, 
as is suggested by some theories of quantum
gravity \cite{Kif99, Klu99}.

\section{Future projects and summary}

There are many projects to be realized in the near future. AGILE \cite{AGI}, a small satellite to be launched in 2003 by the Italian Space Agency, has sensitivity similar to EGRET and will cover the GeV sky before a large and sensitive satellite, GLAST \cite{GLA}, is launched by NASA in 2006. Ground-based observatories are now in the phase of construction. The southern TeV sky will be covered by CANGAROO-III \cite{CAN} (Japan-Australia) and HESS \cite{HES} (mainly Germany), and the northern sky will be viewed by MAGIC \cite{MAG} (mainly Germany) and VERITAS \cite{VER} (mainly USA). The sensitivity will be improved by an order of magnitude in these future detectors.

In summary, gamma-ray observations reveals the non-thermal universe beyond the conventional thermal universe. In particular, observations of the nuclear component (via neutral pion decay) of gamma-rays are important in relation to the long-standing problem of the origins of cosmic rays. Presently the statistics of identified objects are limited, but more objects and new types of targets (e.g., gamma-ray bursts \cite{Der00}, molecular clouds \cite{Aha00},  star burst galaxies \cite{Pag96},  dark matter annihilation \cite{Ber98}) are waiting to be discovered as sensitivity is improved.

\end{document}